\documentclass[aps,twocolumn,superscriptaddress,showpacs,floatfix]{revtex4-1}
\usepackage{amsfonts,amssymb,latexsym,xspace,epsfig,graphicx,color}
\usepackage{amsmath,enumerate,stmaryrd,xy,stackrel,ulem}

\usepackage{xcolor}

\begin{document}
\title{Phase transition in urban agglomeration and segregation}
\author{Taksu Cheon}
\affiliation{Laboratory of Physics, Kochi University of Technology, Tosa Yamada, Kochi 782-8502, Japan}
\author{Michikazu Kobayashi}
\affiliation{Laboratory of Physics, Kochi University of Technology, Tosa Yamada, Kochi 782-8502, Japan}

\date{\today}

\begin{abstract}
A model of the urban agglomeration and segregation is formulated, in which two types of agents move around on the square-lattice aligned cells.  The model is shown to exhibit, when the density of agents are varied as the control parameter, various phase transitions representing appearance of urban aggregation, segregation and social disorder.
\end{abstract}

\maketitle

\section{Introduction}
In social sciences, as in daily conversation on social matters, 
we often encounter the talk about sudden changes between different ''phases''.   However, there has never been a study to show the existence of phase transition in rigorous mathematical sense, as found in models of physical phenomena.  Here we develop a model representing the habitation pattern of agents, which indeed displays phase transition. The model amounts to be an extension of Schelling model of urban segregation, which has been studied in economics in relation to the actual human habitation, but has not received much attention on its mathematical property.

We make two fundamental assumptions on the habitation patterns of people made up from two heterogenous types: the self-aggregation and the xenophobia.  The first refers to the propensity of similar people getting together and clustered, while the second refers to the natural tendency of different people avoiding each other.   We construct a very simple model of two types of agents residing on lattice that incorporates these two dynamics with conditional random relocation in discrete time-steps.  It turns out that the model has rich phase structures, in which one can observe clear-cut phase transitions both of first and second order.

\section{The model}
Consider two types of agents residing on cells arranged on 2-dim square lattice $(i, j)$, $ i=1... n_x$, $j=1... n_y$ with periodic boundaries.
Each cell can have three states: $x(ij)=0, 1, -1$, respectively signifying the empty cell, the presence of a type-1 agent, and the presence of a type-2 agent.

We start from random initial state with densities $\rho_1$ and $\rho_2$ for respective types. We use total density $\rho$ and asymmetry $\beta$ defined by
$\rho = \rho_1 + \rho_2$ and $\beta =(\rho_1 - \rho_2)/(\rho_1 + \rho_2)$.  Consider a cell inhabited by an agent, and define the numbers $m_{us}$ and $m_{th}$ for that cell as
the number of own type (“us”), and the number of other type (“them”) among its neighboring eight cells, respectively.  

The time evolution of the system is determined by an update rule, specified by two threshold numbers $N_S$ and $N_A$, which dictate that an agent {\bf changes its location} if 
\begin{eqnarray}
(S): n_{th} \geqslant N_S \qquad {\it and}\qquad  (A): n_{us}<N_A.  
\end{eqnarray}
We can limit the range of each threshold number to be integer in the rage $9  \geqslant N_S \geqslant 1$ and $8 \geqslant N_A \geqslant 0$.  The update rule can be rephrased as each agent {\bf stays in its location} when conditions
\begin{eqnarray}
\!\!\!\!
(\bar{S}) : n_{th} < N_S \qquad {\it or} \qquad (\bar{A}) :  n_{us}\geqslant N_A   
\end{eqnarray}
are met.  The first condition represents the ''xenophobic'' tendency of both types, and the second the tendency to cluster among themselves.

\section{Pure self-aggregation limit : $N_S=0$}

We first consider the case of $N_S=0$, in which each agent relocates if its eight neighboring cells contain less than $N_A$ agents of its own type.  

 \begin{figure}[htbp] 
   \centering
   \includegraphics[width=7cm]{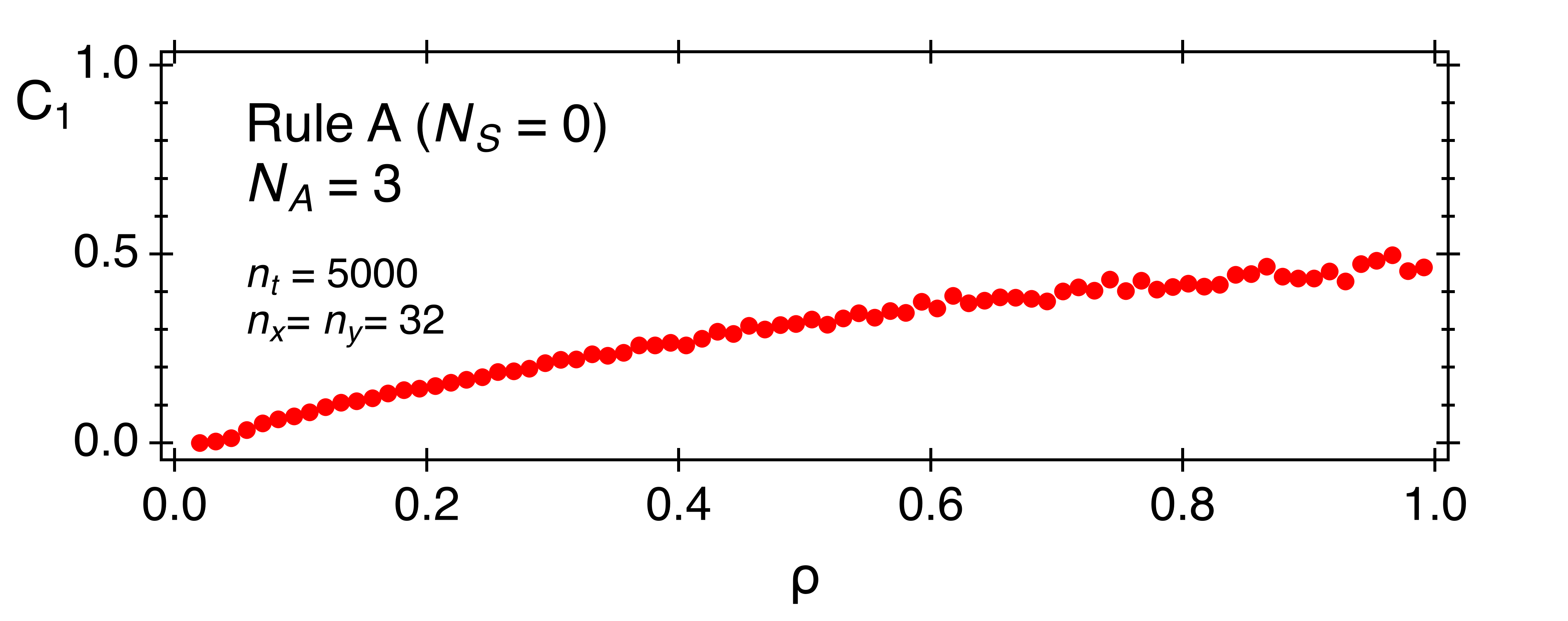} 
   \includegraphics[width=7cm]{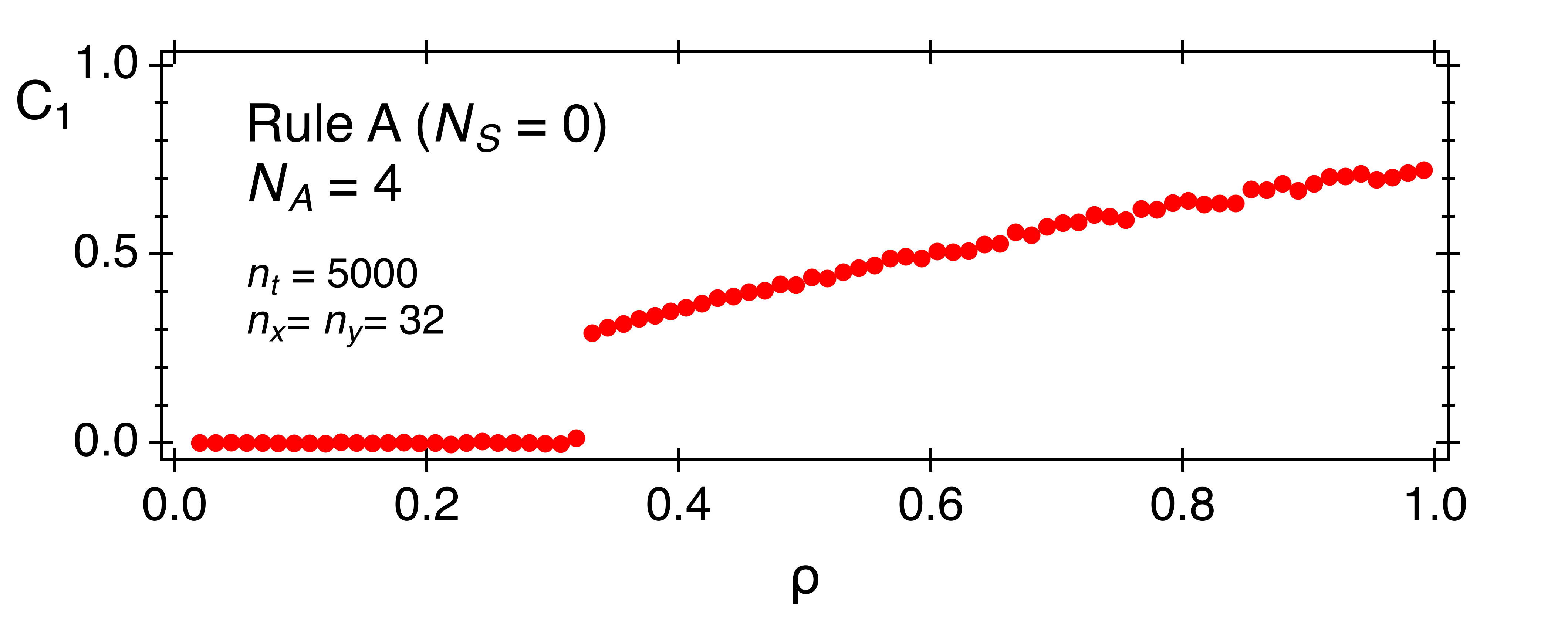}
   \includegraphics[width=7cm]{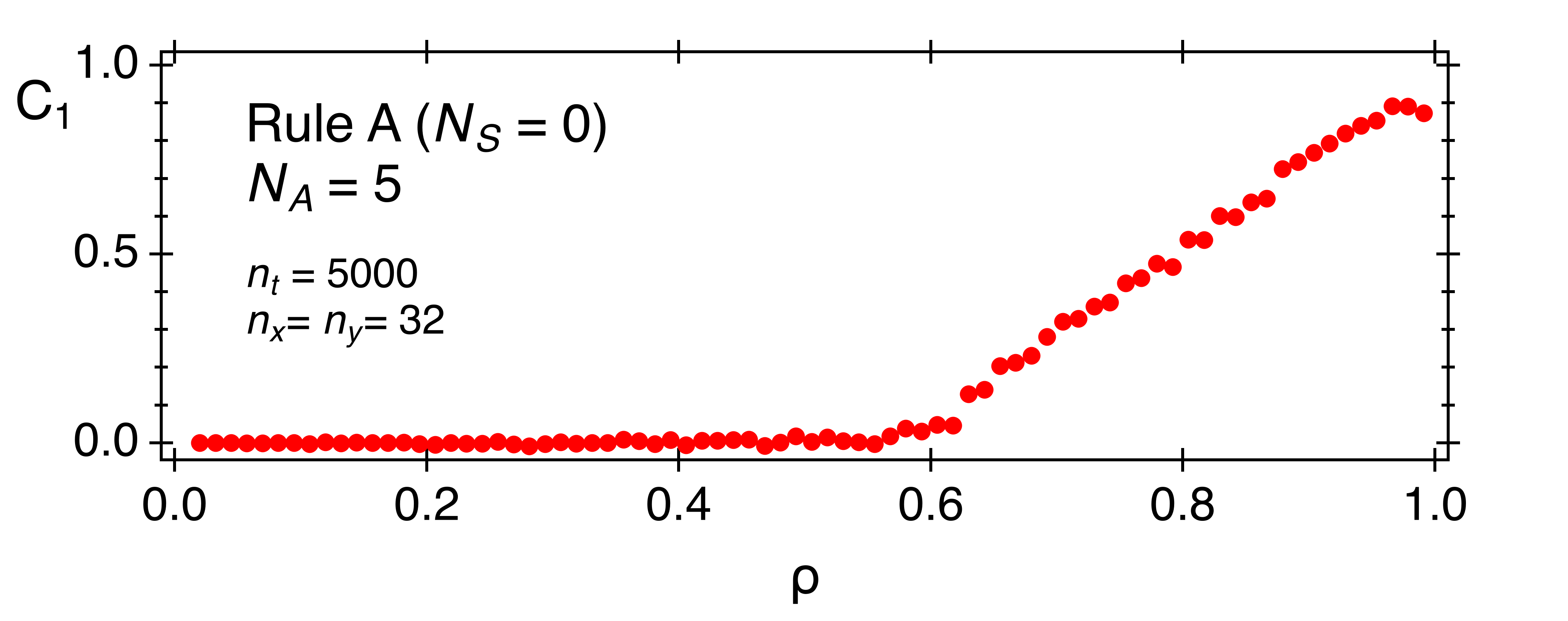}  
   \caption{Nearest neighbor correlation as a function of agent density for the case of  $N_s=0$ and $N_A=3$ (top), $N_S=0$ and $N_A=4$ (middle), $N_s=0$ and $N_A=5$ (bottom). }
   \label{fig1}
\end{figure}

We calculate the neighboring correlation $C_1$ which is defined by
\begin{eqnarray}
C_1 = \frac{\sum_{(i, j)}\sum_{(k, l)={\rm next\ neighbers\ of} (i, j)} x_{i,j} x_{k, l}}
{\sum_{(i, j)}\sum_{(k, l)={\rm next\ neighbers\ of} (i, j)} }
\end{eqnarray}
as functions of $\rho$.
  
We show the results of numerical calculation of $C_1(\rho)$ in Figure \ref{fig1}.  We only consider symmetric case, $\beta=0$ in this work.  The threshold parameter is set to be $N_A=3$, $4$, and $5$.  In the calculation, we have used the lattice size $n_x = n_y = 32$, and the maximum time step $n_t=5000$.   

For $N_A=3$, we find smooth growth of agglomeration as we increase $\rho$.   A discontinuous $C_1(\rho)$ for $N_A=4$ around $\rho=0.3$ indicating the sudden emergence of clustering is observed for this case.  

\begin{figure}[htbp] 
   \centering
   \includegraphics[width=1.5cm]{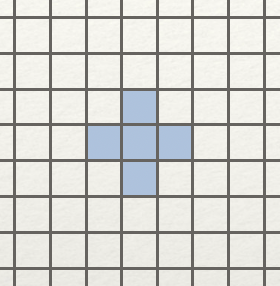} \qquad\qquad
   \includegraphics[width=1.5cm]{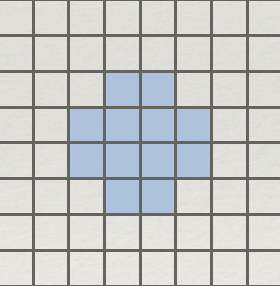} 
   \caption{Minimal stable cluster for $N_A=3$ (left) and $N_A=4$ (right).}
   \label{fig20}
\end{figure}

The reason of this first-order phase transition-like feature can be understood as follows.  At the low density, the clustering of agent starts with the formation of stable minimal cluster which is made up of twelve agents (Figure \ref{fig20} right).  

\begin{figure}[htbp] 
   \centering
   \includegraphics[width=8cm]{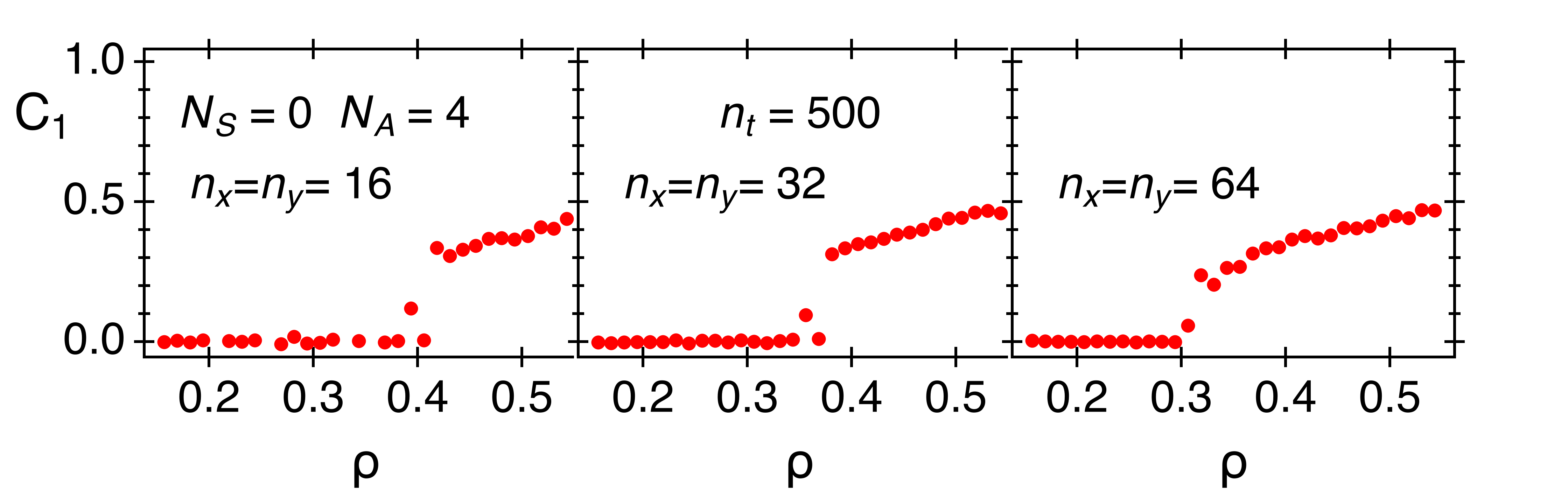} 
   \includegraphics[width=8cm]{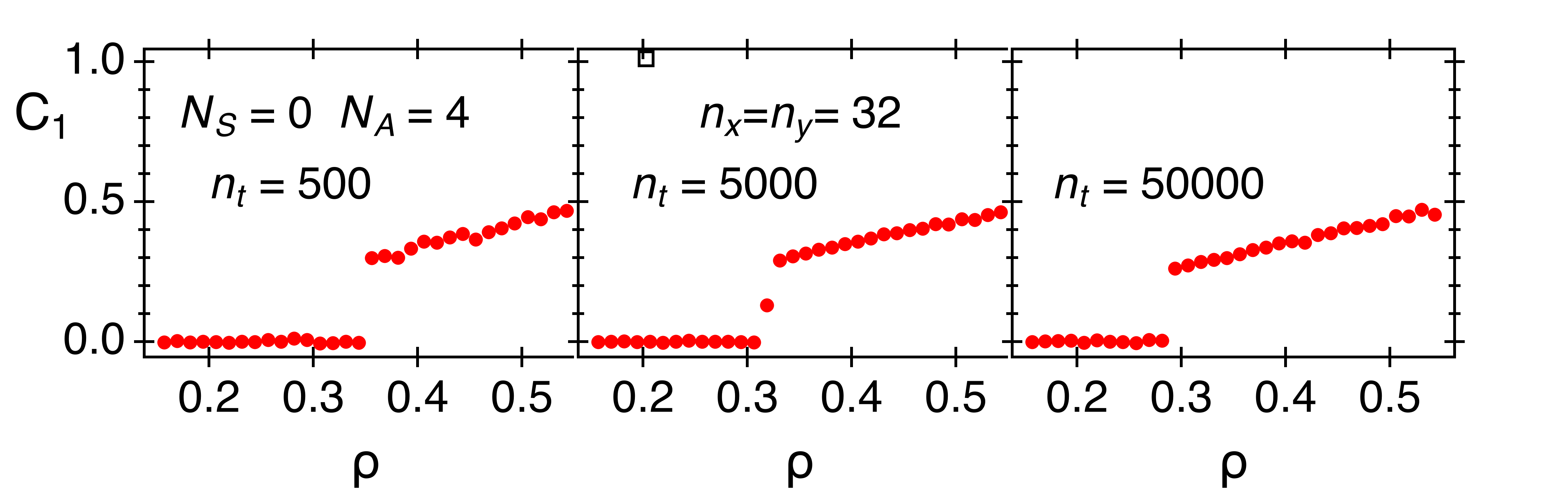}
   \caption{Nearest neighbor correlation as a function of agent density for the case of  $N_S=0$ and $N_A=4$ with changing tempporal cut-ff(top), and lattivce size  (bottom). }
  \label{fig54}
\end{figure}

\begin{figure}[htbp] 
   \centering
   \includegraphics[width=8cm]{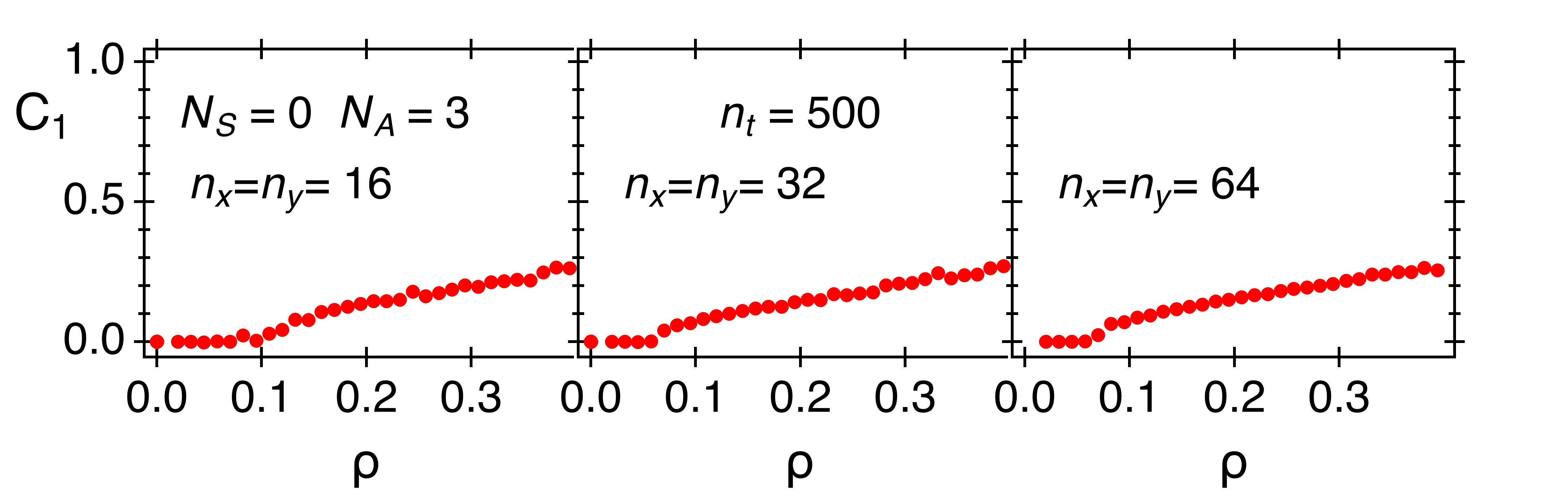} 
   \includegraphics[width=8cm]{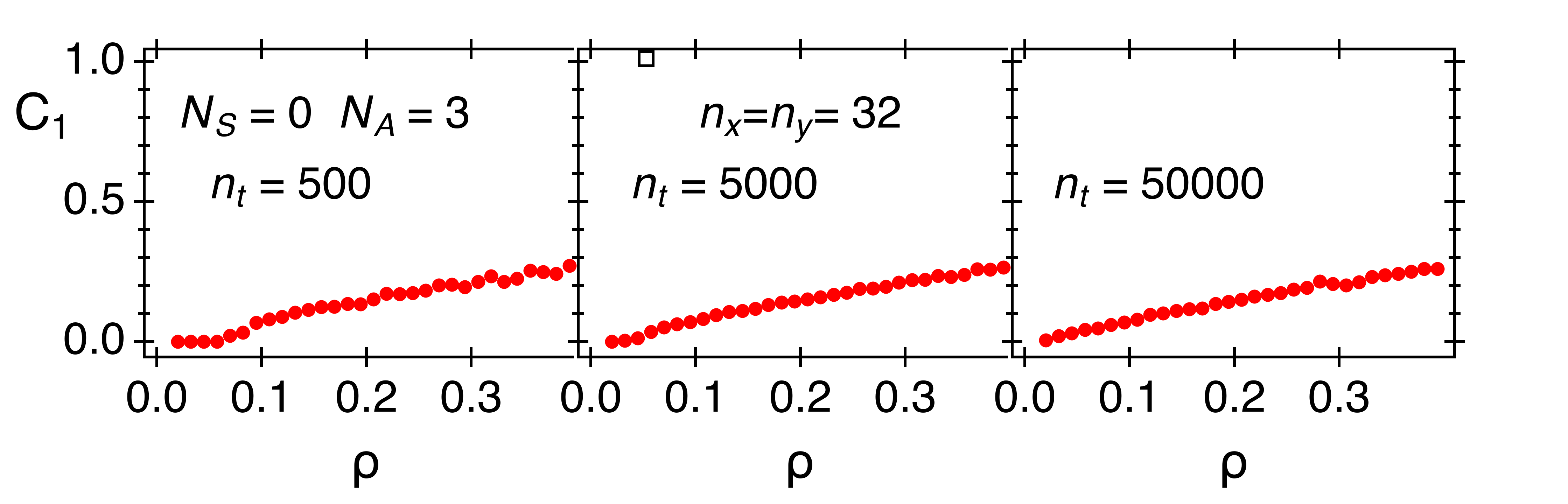}
   \caption{Nearest neighbor correlation as a function of agent density for the case of  $N_S=0$ and $N_A=3$ with changing tempporal cut-ff(top), and lattivce size  (bottom). }
   \label{fig53}
\end{figure}

We can make a rough estimate of the probability of this formation as $n_x n_y \rho^{12}$, whose inverse should give the necessary time step  $n_t$.   This gives us an estimate for the critical density $\rho_{cr} = (n_t n_x n_y)^{-\frac{1}{12}}$ which goes along with our numerical calculation reasonably well.  Thus this discontinuity in $C_1(\rho)$ is a pseudo phase transition\cite{IS71} akin to glass transition which arise as the finite lattice size and finite time effect, rather than a true phase transition that persists to $n_x = n_y$ $\to \infty$ and $n_t \to \infty$ limit.  This estimate  is shown to be reasonable valid to predict the $n_t$ and $n_x=n_y$ dependence of $\rho_{cr}$ shown in Figure \ref{fig54}.

We note that it is possible to create a model that has genuine first-order phase transition by introducing a time scale, say, by way of percolation probability $p_p = 1/n_t$ into the model.  

An analogous discontinuity in $C_1(\rho)$ should be found for $N_A=3$ case also, with the minimum stable cluster given by aggromeration of five agents shown in Figure \ref{fig20} left.  This gives us an estimate $\rho_{cr} = (n_t n_x n_y)^{-\frac{1}{5}}$, which is corroborated by our calculation of $C_1(\rho)$ with various $n_t$ and $n_x=n_y$ shown in Figure \ref{fig53}.
%

\begin{figure}[htbp] 
   \centering
   \includegraphics[width=8cm]{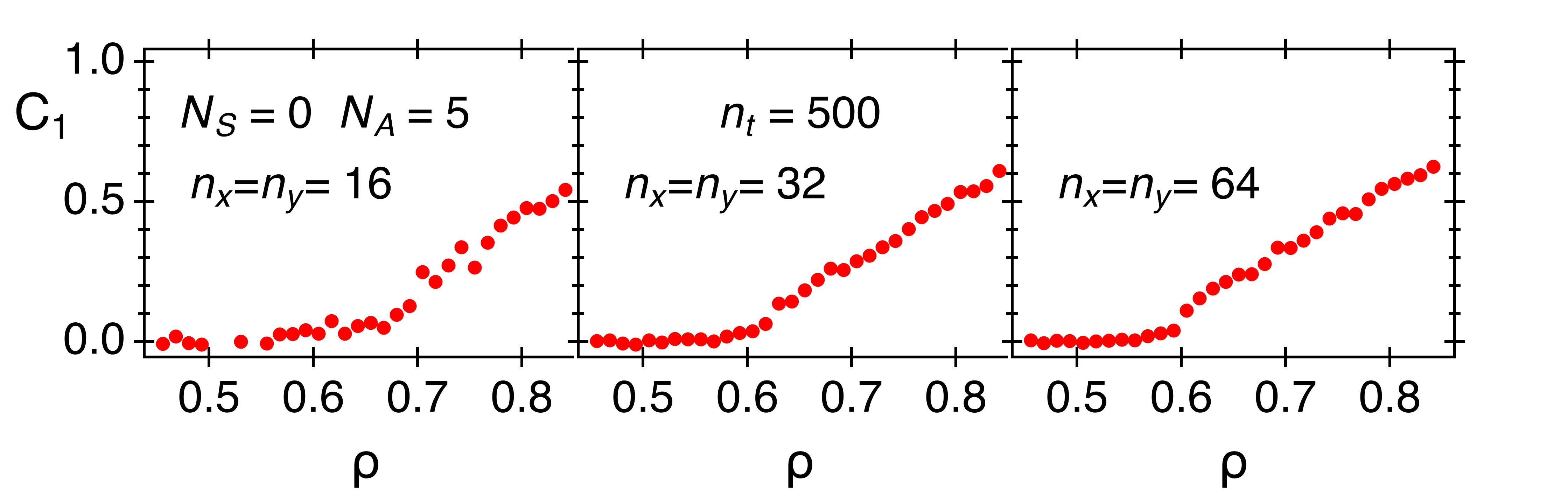} 
   \includegraphics[width=8cm]{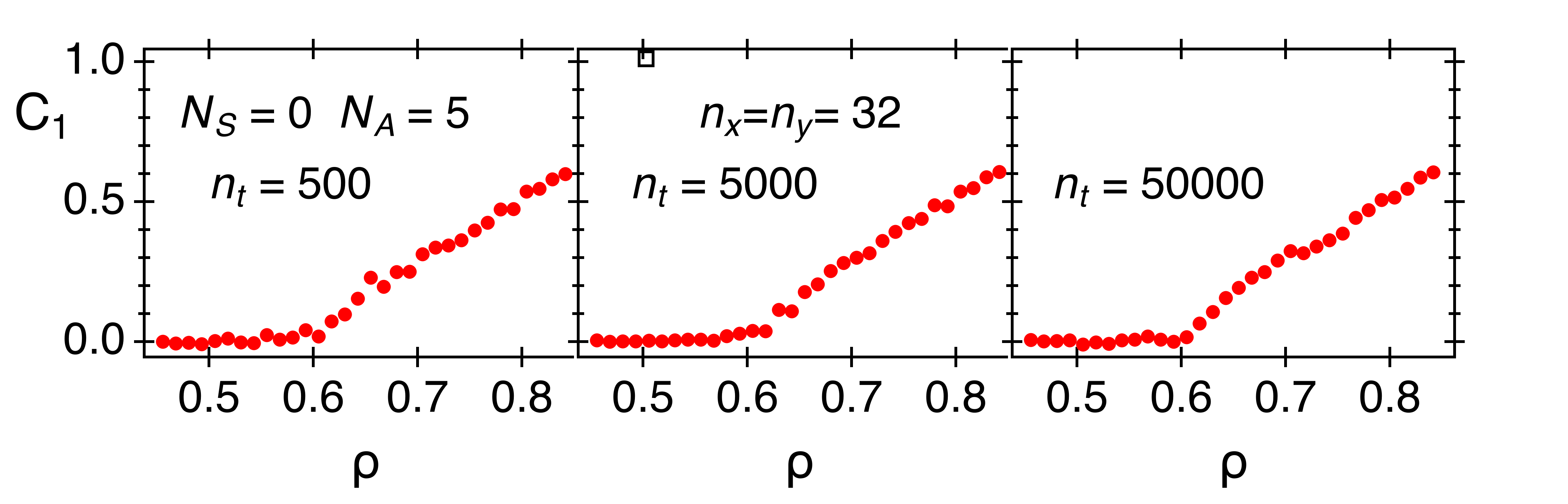}
   \caption{Nearest neighbor correlation as a function of agent density for the case of  $N_S=0$ and $N_A=5$ with changing tempporal cut-ff(top), and lattivce size  (bottom). }
   \label{fig51}
\end{figure}

For $N_a=5$ case, we observe the appearance of grid-like structure at $\rho=0.5$ which alternate empty and filled raws at each time steps.  After further increase of the density, we encounter the second-order phase transition at around $\rho=6$ beyond which clustering of each type separated by grid-structure occurs.  The stability of the calculated critical point with respective to the variation of $n_t$ and $n_x=n_y$ confirmes that this is a genuine phase transition  which persist at continuous limit (Figure \ref{fig51}).

 \begin{figure}[htbp] 
   \centering
 \includegraphics[width=8cm]{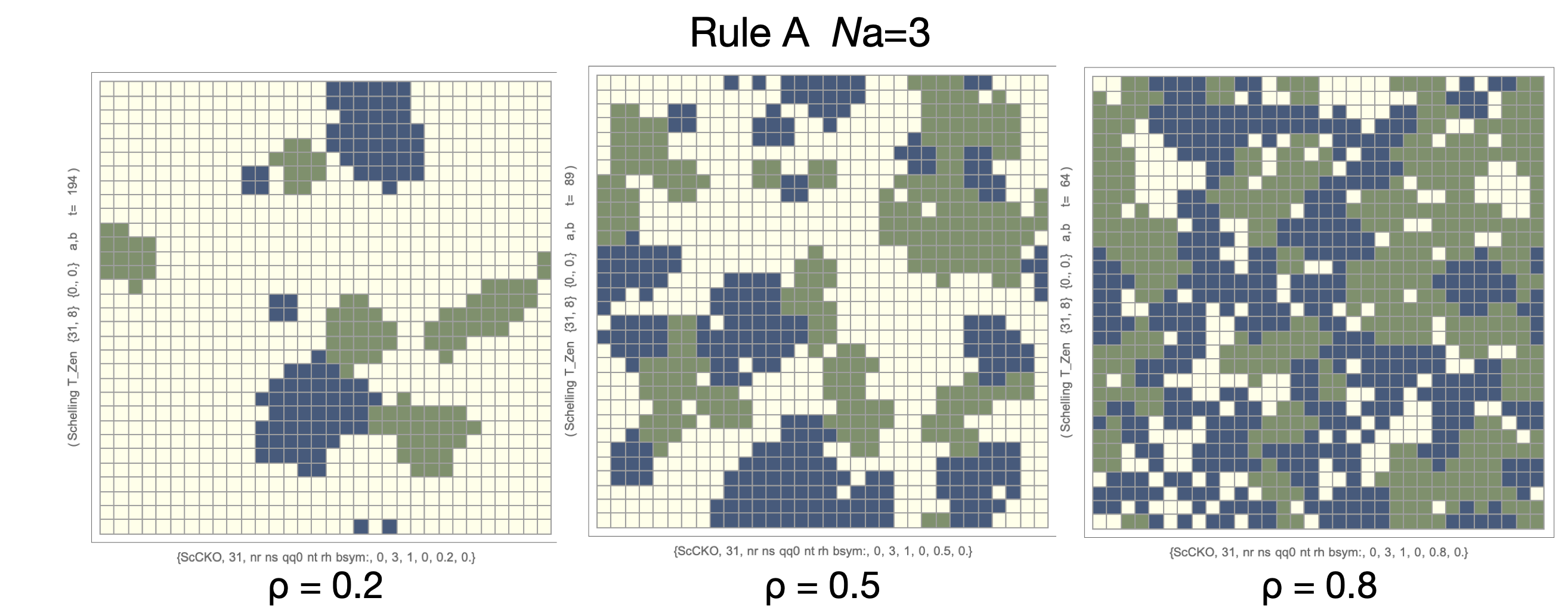} 
  \centering
 \includegraphics[width=8cm]{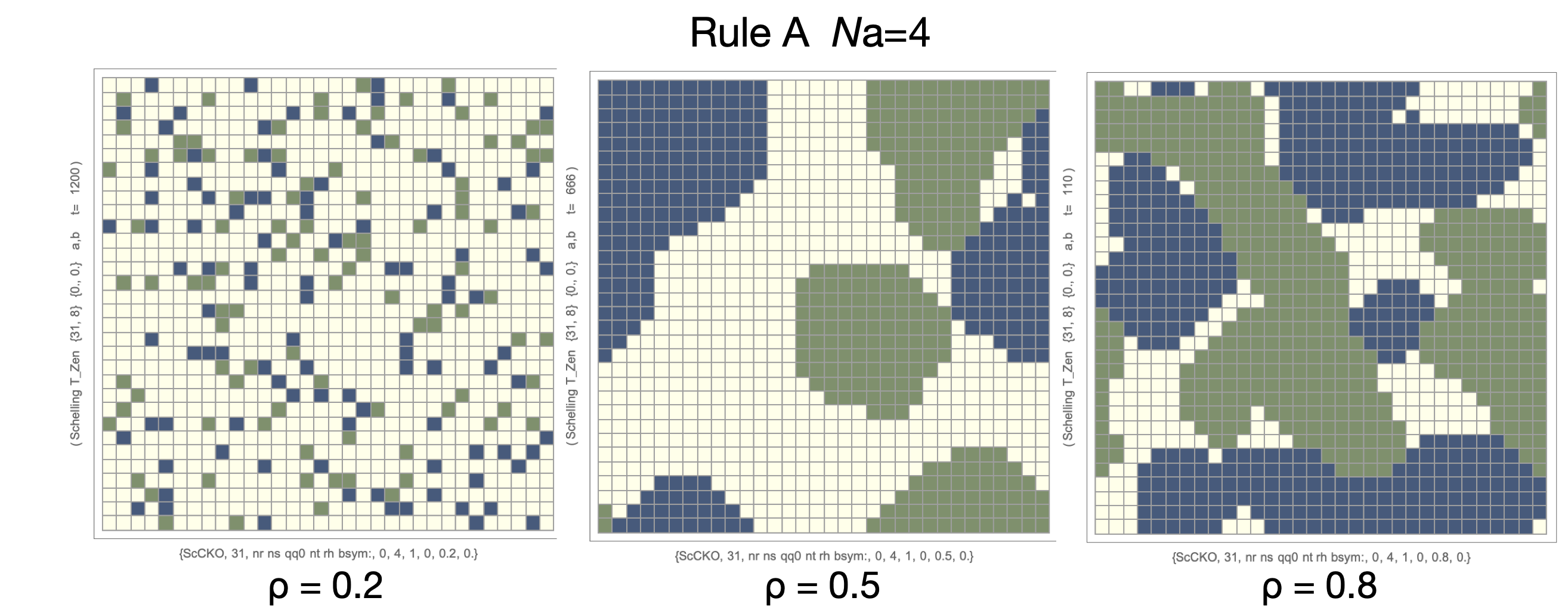} 
  \centering
 \includegraphics[width=8cm]{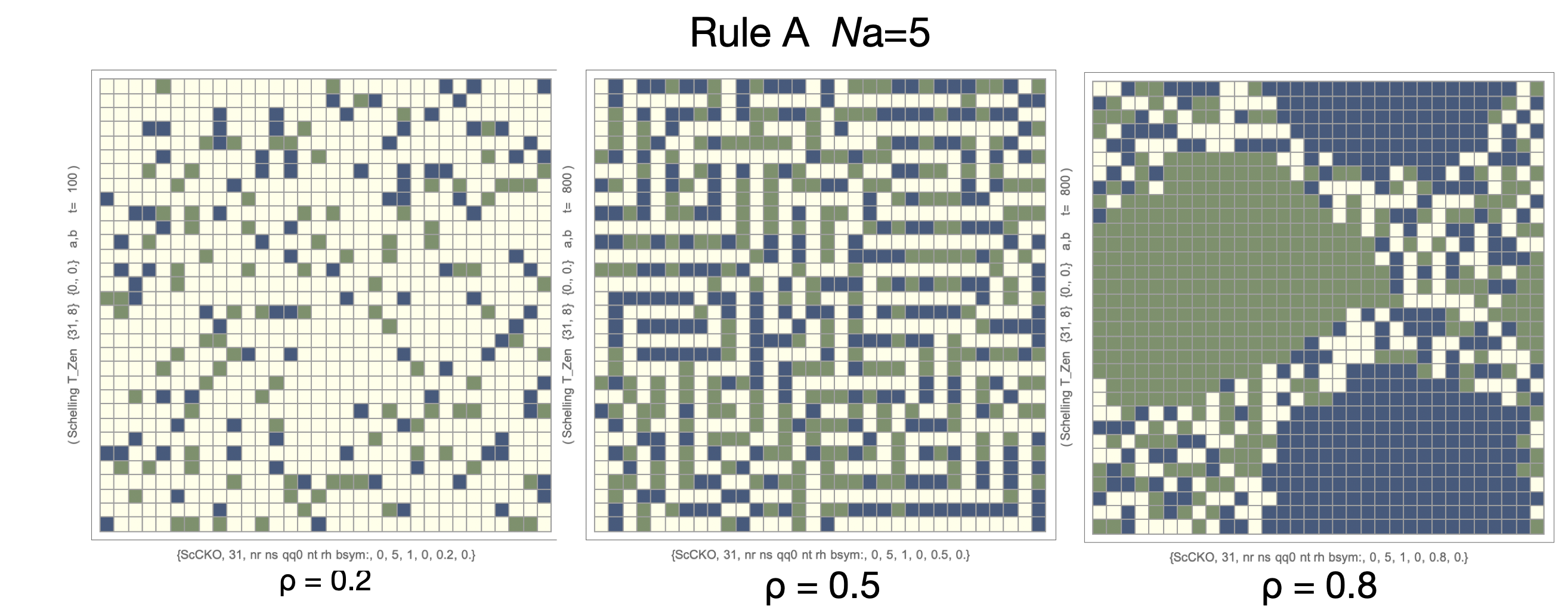} 
 \caption{Spacial pattern of agent configuration at time-step large enough to have stable final state.  The model parameters $N_A=3$ (top), $N_A=4$ (middle),$N_A=4$ (bottom), with common number $N_S=0$.}
   \label{fig12}
\end{figure}

A typical spatial aggregation patterns are shown in Figure \ref{fig12}, which depicts the spacial configuration of fully developed states of the system for 
$N_A=3$, $4$ and $5$, from top to bottom.  
For $N_A=3$, all states shown are stable final states.  For $N_A=4$, the state shown for $\rho=0.2$ is a snapshot of states with perpetual random motion, while the states for $\rho=0.5$ and $\rho=0.8$ are both stable final states.  $N_A=3$, all states shown are stable final states.  For $N_A=5$, all depicted are snapshots of changing meta-stable states. 

\section{Schelling limit, purely xenophobic agents: $N_a=9$} 

We now consider the case of $N_a=9$, that is, each agent relocates if its eight neighbor contains more than or equal to  $N_s$ agents of the other type.  This limit corresponds to a generalization of selebrated Schelling model of racial segregation \cite{Sch71}.
We show the numerical results of $C_1(\rho)$.

 \begin{figure}[htbp] 
   \centering
   \includegraphics[width=7cm]{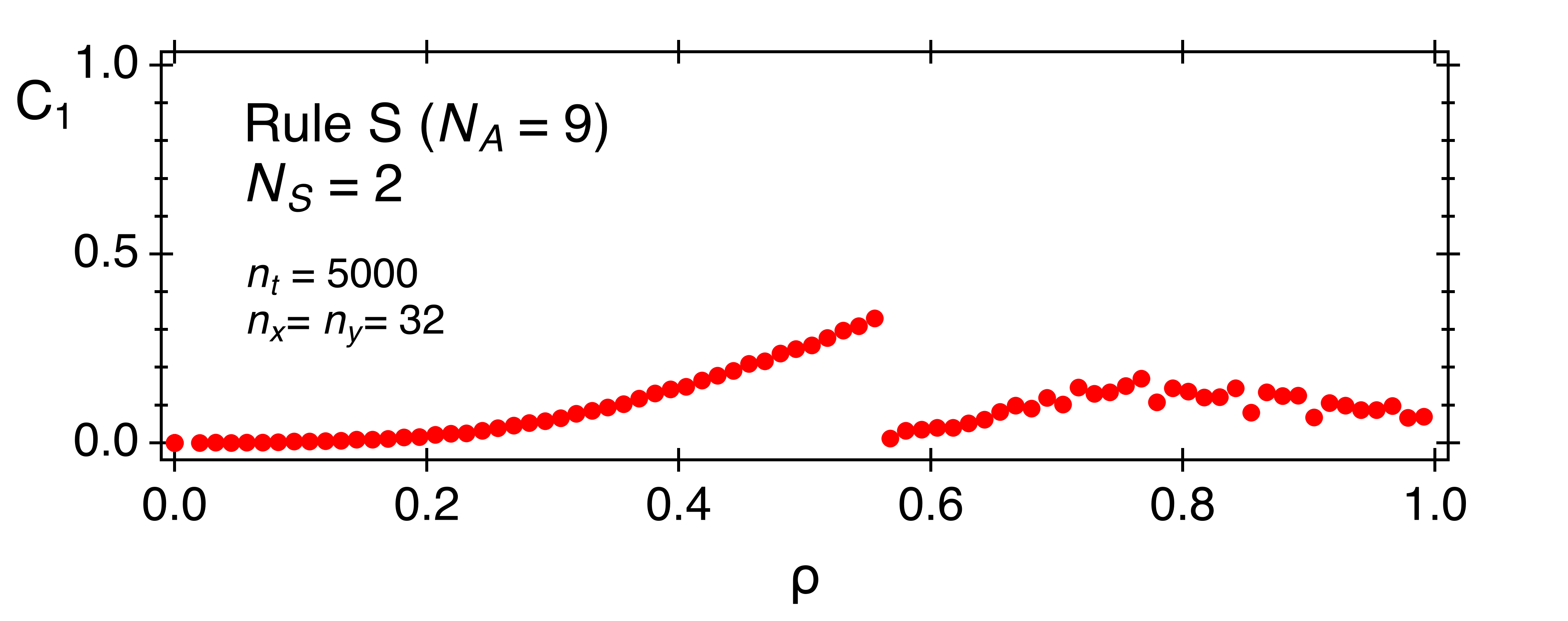} 
   \includegraphics[width=7cm]{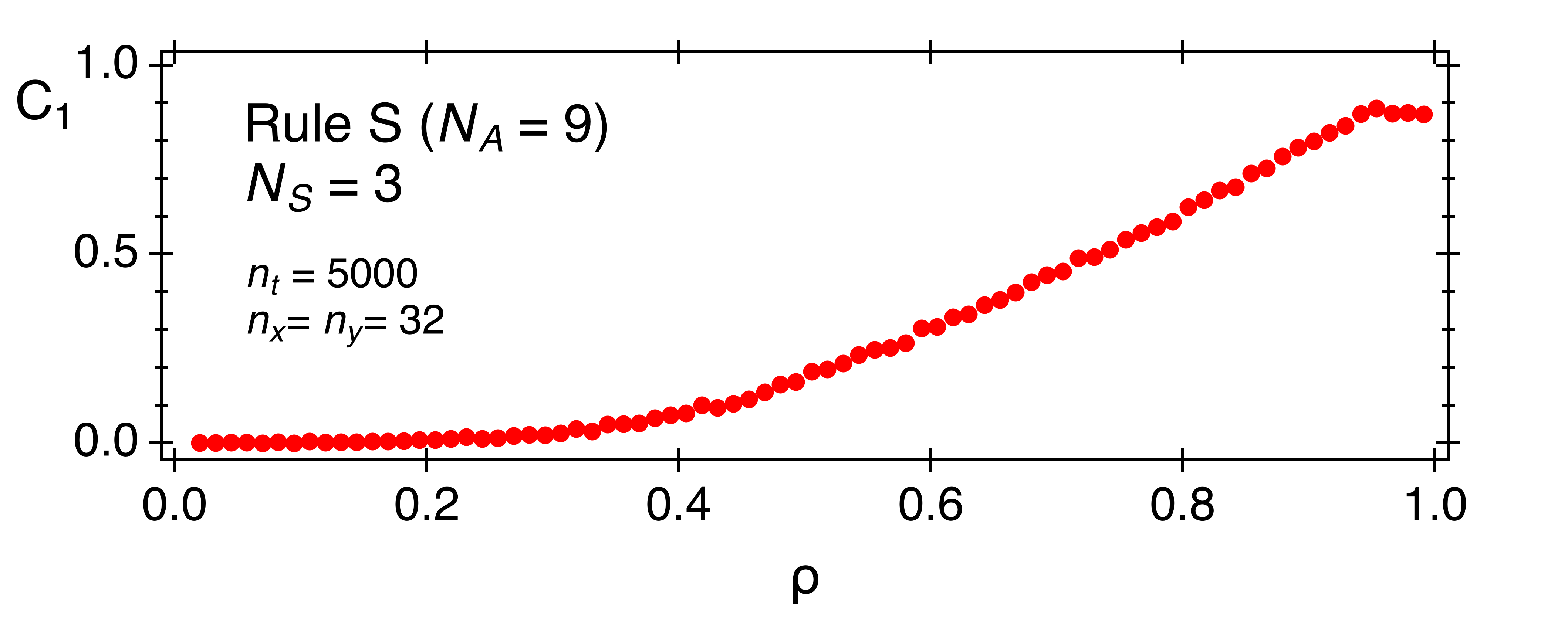} 
   \caption{Nearest neighbor correlation as a function of agent density for the case of  $N_S=2$ and  (top), $N_S=3$  (bottom) with common value and $N_A=9$.}
   \label{fig4}
\end{figure}

The threshold parameter is set to be $N_s=2$, $3$, and $4$. We have used $n_x = n_y = 32$, and time steps $n_t=5000$.  We choose $\beta=0$ and the system is set to be symmetric with hrespect to type one and two.

We show the neighboring correlation $C_1$ as functions of the density $\rho$ in Figure \ref{fig4}.
Both for $N_s=2$ and $N_s=3$ cases, the states shown for $\rho=0.2$ is a snapshot of  perpetual random motion, while the states for $\rho=0.5$ and $\rho=0.8$ are both stable final states.

\begin{figure}[htbp] 
   \centering
   \includegraphics[width=8cm]{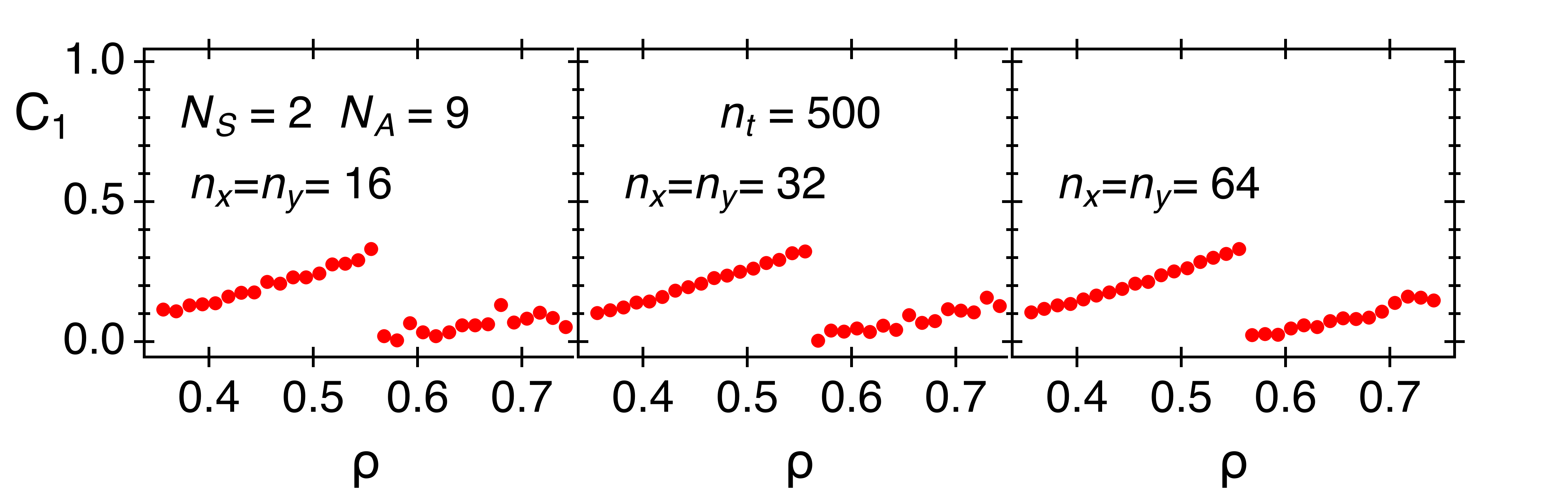} 
   \includegraphics[width=8cm]{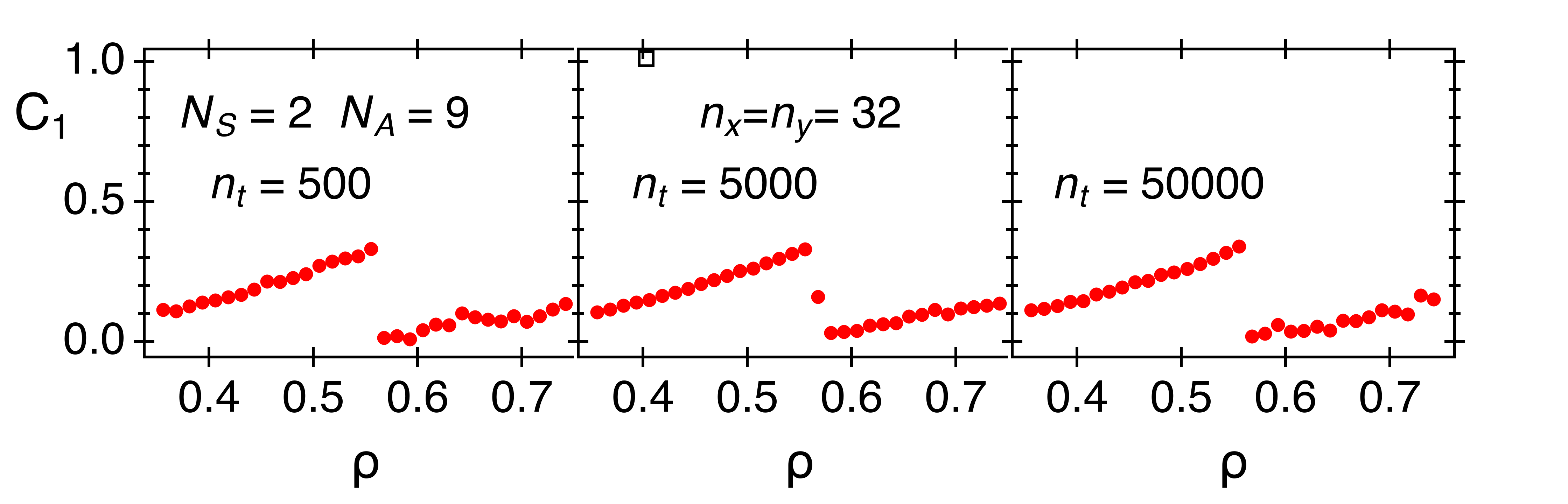}
   \caption{Nearest neighbor correlation as a function of agent density for the case of  $N_s=2$ and $N_a=9$ with changing temporal cut-ff(top), and lattice size  (bottom). }
   \label{fig52}
\end{figure}

Both for $N_s=2$ and $N_s=3$, we find smooth emergence of segregational clustering as we increase $\rho$.  While for $N_s=3$, the trend last to the fully packed density $\rho=1$, a remarkable feature is observed foe the case of $N_s=2$, in which the existence of the first-order phase transition is clearly seen.  The stability of the calculated critical point with respective to the variation of $n_t$ and $n_x=n_y$ confirmes that this is a genuine phase transition  which persist at continuous limit (Figure \ref{fig52}).

 \begin{figure}[htbp] 
   \centering
   \includegraphics[width=8cm]{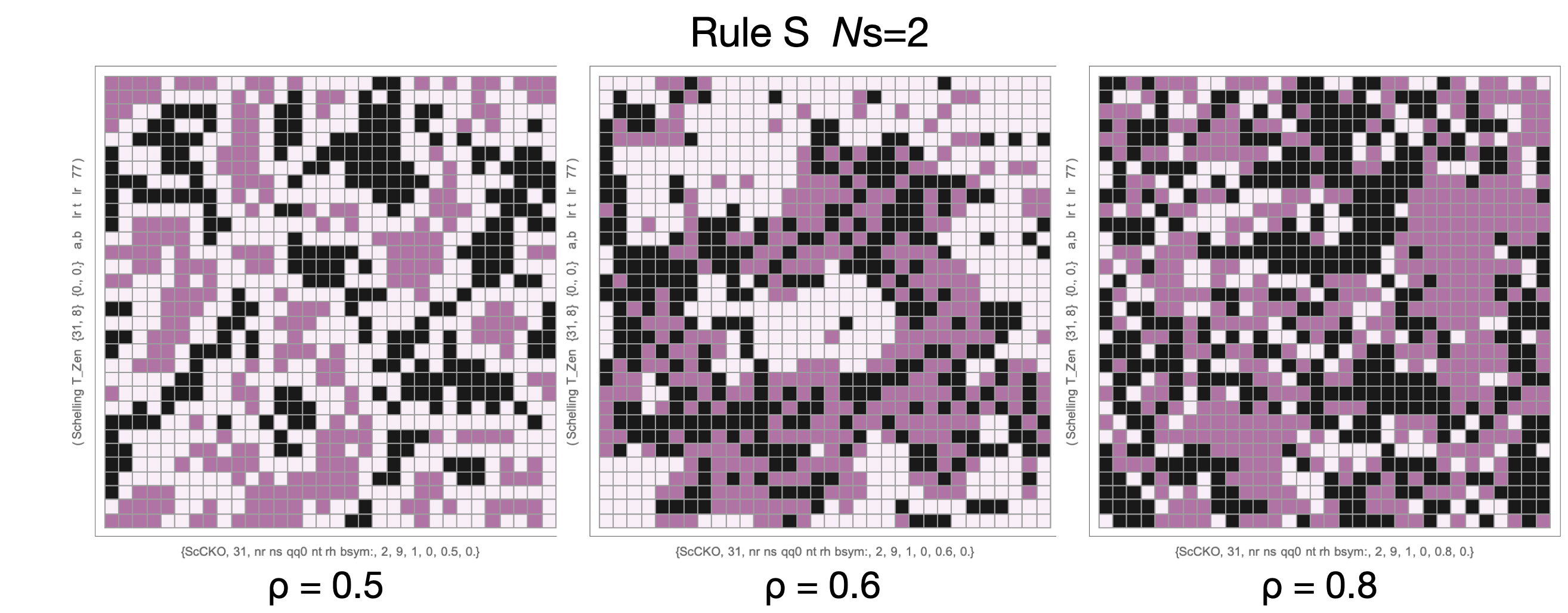} 
    \centering
   \includegraphics[width=8cm]{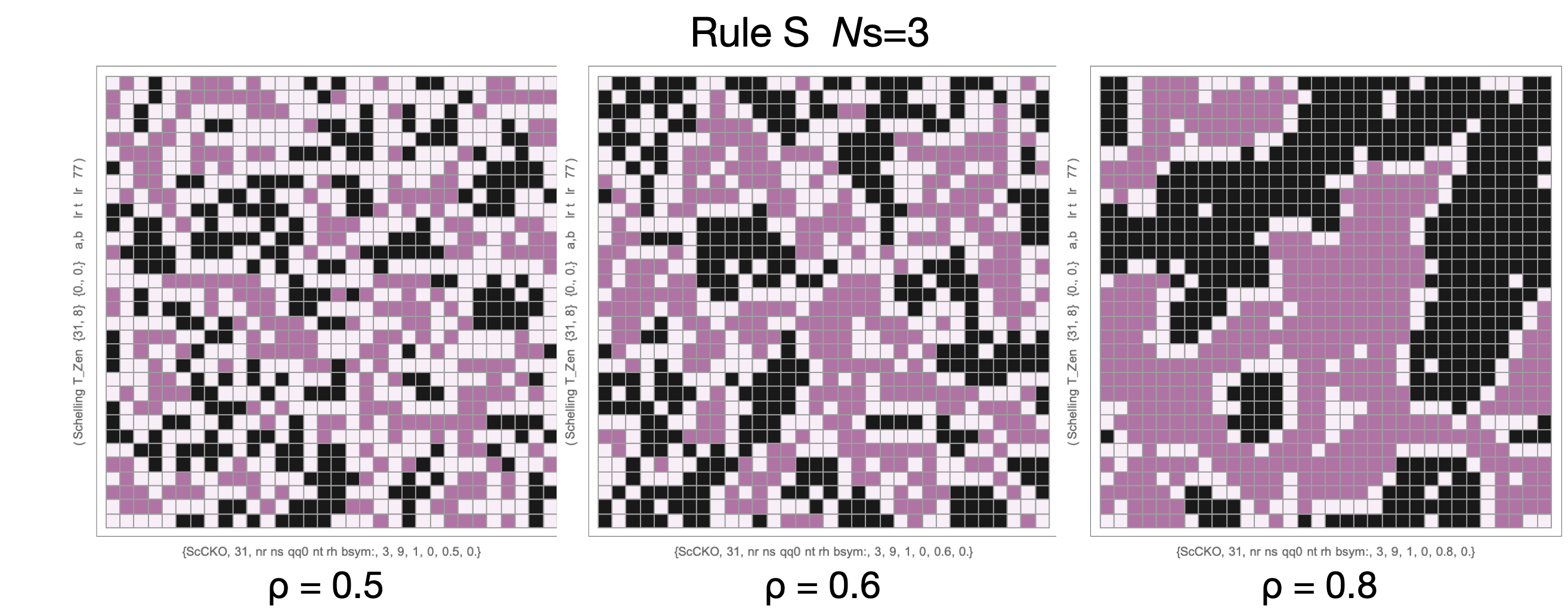} 
   \caption{Spacial pattern of agent configuration at time-step large enough to have stable final state.  The model parameters are $N_s=2$, $N_a=9$ (top), and $N_s=3$, $N_a=9$ (bottom).}
   \label{fig7}
\end{figure}

\begin{figure}[htbp] 
   \centering
   \includegraphics[width=8cm]{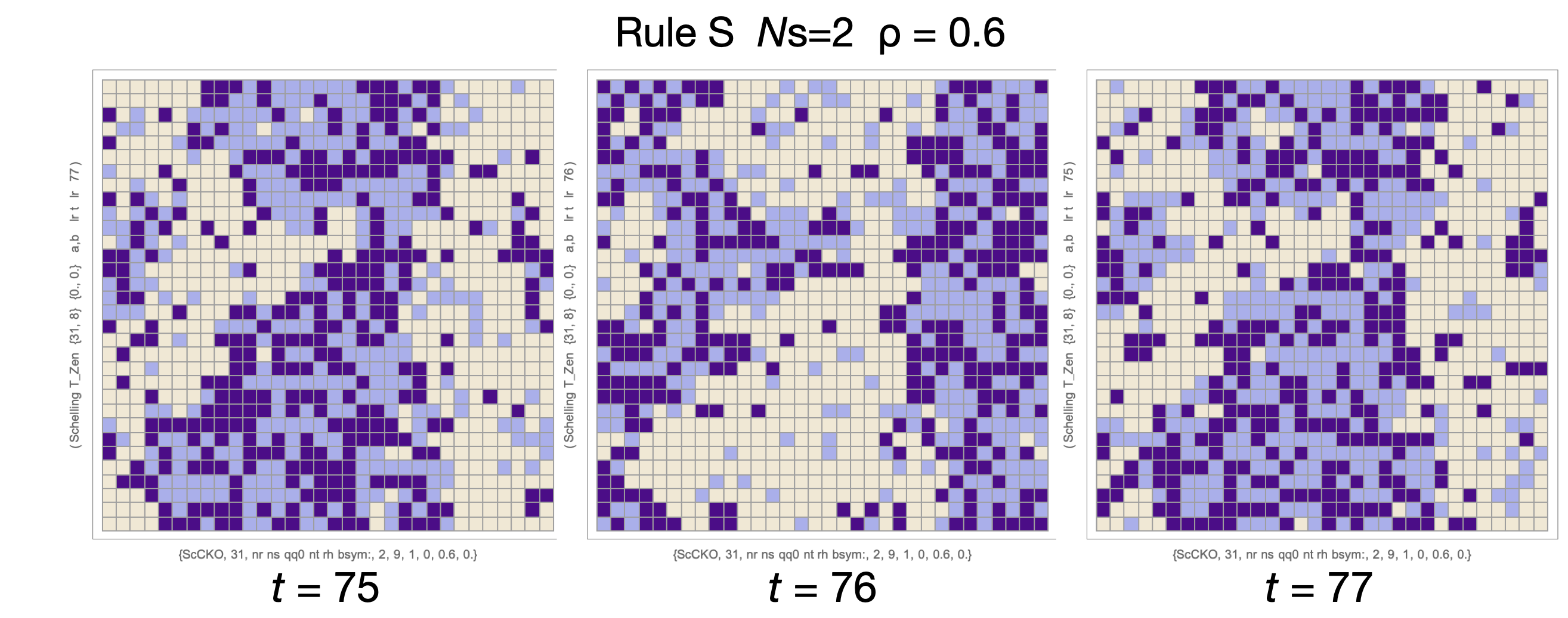} 
   \caption{Temporal evolution of spacial pattern of agent configuration for $\rho=0.6$, just above the ''social unrest'' transition.}
   \label{fig9}
\end{figure}

We also show the spatial patters of agents filling the torus in Figure \ref{fig7}.  The origin of the discontinuous transition for $N_s=2$ is the inability of the system to have stabel final state indicating the social unrest \cite{BR12}.  The situation can be  understood by inspecting the temporal variation of spatial pattern of agents depicted in Figure \ref{fig9}.

\section{General case}
\begin{figure}[htbp] 
   \centering
   \includegraphics[width=8cm]{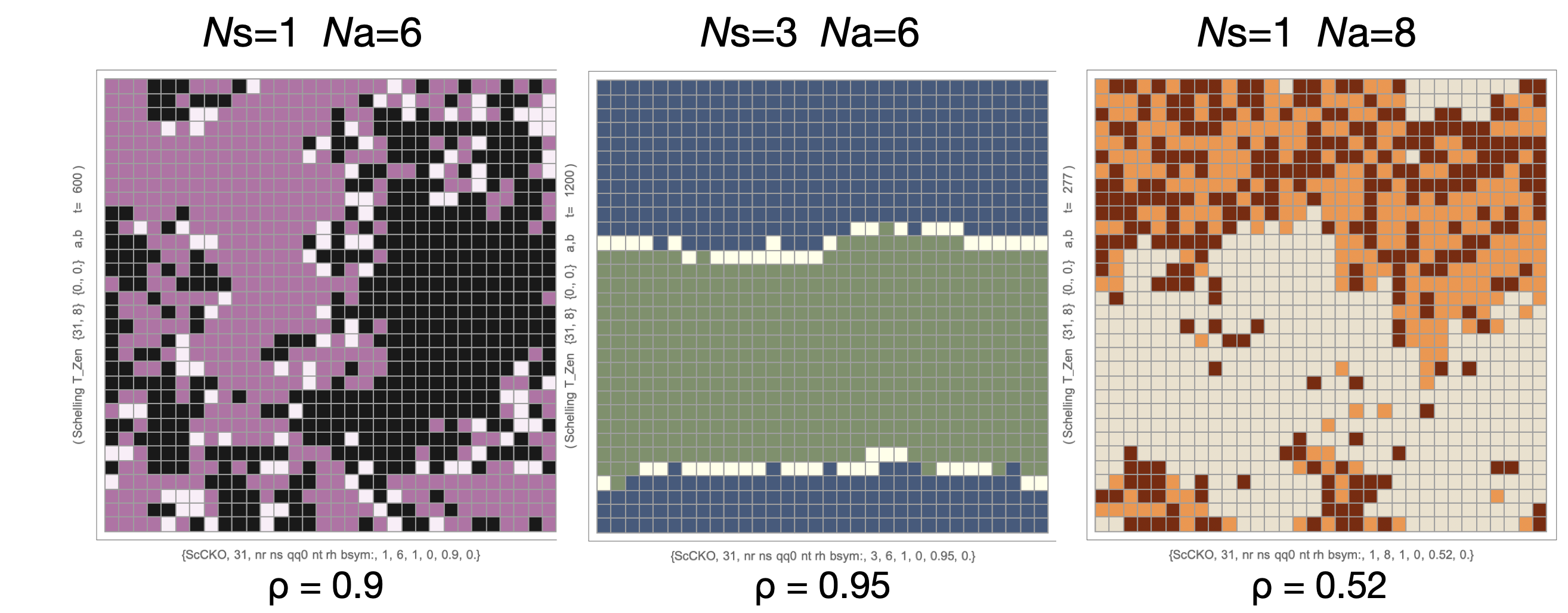} 
   \caption{Spacial patterns of agent configurations  for models with various threshold parameters $N_s$ and $N_a$. }
   \label{fig13}
\end{figure}

For the general case of $N_s \ne 0$ or $N_a \ne 9$, we find all types of phase transitions similar to above limiting cases.  The habitation patterns obtained from general model is quite varied as can be expected, from which we only show several striking examples in Figure \ref{fig13}.

\section{Summary and conclusion}

Intriguing phase transitions are found in urban agglomeration model that comprises self-aggregating and xenophobic dynamics.  

We find it intriguing that, for the case of $N_A=9$, which amounts to the Schelling model, we not only find stable segregation but also social unrest depending on the parameter $N_s$ which signifies the degree of animosity among different types of agents.

The simple model developed here has a strong potential for further extensions.  The three-agent model, an immediate and obvious extension, seem to offer a promising ground for richer features with direct relevance to observed phenomena in actual societies.   
We note that this model might offer simple alternative to celebrated Turing mechanism \cite{TU52} for pattern generation.


\begin{thebibliography}{9}

\bibitem{Sch71}
T. C. Schelling, . "Dynamic models of segregation". J. Math. Soc.  {\bf 1}  (1971)  143--186.

\bibitem{IS71}
Y. Imry and D. J. Scalapino, "Pseudo-first-order phase transitions in one dimension". Phys. Rev. A  {\bf 9}  (1974)  1673--1674.

\bibitem{BR12}
D. Braha, "Global Civil Unrest: Contagion, Self-Organization, and Prediction". PLOS ONE {\bf 7} (2012) e48596 (9pp).
 
\bibitem{TU52}
A. Turing,   "The chemical basis of morphogenesis", Phil.Trans. Roy. Soc. London B 237 (1952) 37--72.

\end{thebibliography}
\end{document}